\documentclass[12pt]{article}
\usepackage{amsmath,amsfonts,amssymb}

\begin{document}
\thispagestyle{empty}

\def\theequation{\arabic{section}.\arabic{equation}}
\def\a{\alpha}
\def\b{\beta}
\def\g{\gamma}
\def\d{\delta}
\def\dd{\rm d}
\def\e{\epsilon}
\def\ve{\varepsilon}
\def\z{\zeta}
\def\B{\mbox{\bf B}}

\newcommand{\h}{\hspace{0.5cm}}

\begin{titlepage}
\vspace*{1.cm}
\renewcommand{\thefootnote}{\fnsymbol{footnote}}
\begin{center}
{\Large \bf Some three-point correlation functions in the
$\eta$-deformed $AdS_5\times S^5$}
\end{center}
\vskip 1.2cm \centerline{\bf Plamen Bozhilov} \vskip 0.6cm
\centerline{\sl Institute for Nuclear Research and Nuclear Energy}
\centerline{\sl Bulgarian Academy of Sciences} \centerline{\sl
1784 Sofia, Bulgaria}

\centerline{\tt bozhilov@inrne.bas.bg, bozhilov.p@gmail.com}

\vskip 20mm

\baselineskip 18pt

\begin{center}
{\bf Abstract}
\end{center}
\h We compute some normalized structure constants in the
$\eta$-deformed $AdS_5\times S^5$ in the framework of the
semiclassical approach. This is done for the cases when the
``heavy'' string states are finite-size giant magnons carrying one
angular momentum and for three different choices of the ``light''
state: primary scalar operators, dilaton operator with nonzero
momentum, singlet scalar operators on higher string levels. Since
the dual field theory is still unknown, the results obtained here
must be considered as conjectures or as predictions from the
string theory side.



\end{titlepage}
%\end{quote}
%\vskip 1cm \centerline{\today}
\newpage
%\baselineskip 18pt

%%%%%%%%%%%%%%%%%%%%%%%%%%%%%%%%%%%%%%%%%%%%%%%%%
\def\nn{\nonumber}
%%%%%%%%%%%%%%%%%%%%%%%%%%%%%%%%%%%%%%%%%%%%%%%%%
\def\tr{{\rm tr}\,}
\def\p{\partial}
\newcommand{\bea}{\begin{eqnarray}}
\newcommand{\eea}{\end{eqnarray}}
\newcommand{\bde}{{\bf e}}
\renewcommand{\thefootnote}{\fnsymbol{footnote}}
\newcommand{\be}{\begin{equation}}
\newcommand{\ee}{\end{equation}}
%\newcommand{\h}{\hspace{0.5cm}}
%%%%%%%%%%%%%%%%%%%%%%%%%%%%%%%%%%%%%%%%%%%%%%%%%%%%

\vskip 0cm

\renewcommand{\thefootnote}{\arabic{footnote}}
\setcounter{footnote}{0}

%\setcounter{equation}{0}
%\section{Introduction}

\setcounter{equation}{0}
%%%%%%%%%%%%%5%%%%%%%%%%%%%%%%%%%%%%%%%%%%%%%%%%%%%%%%%%%%%%%%%%%%%%%%%%%%%%%%
\section{Introduction}
%%%%%%%%%%%%%5%%%%%%%%%%%%%%%%%%%%%%%%%%%%%%%%%%%%%%%%%%%%%%%%%%%%%%%%%%%%%%%%
The AdS/CFT duality \cite{AdS/CFT}-\cite{EW98} between string
theories on curved space-times with Anti-de Sitter subspaces and
conformal field theories in different dimensions has been actively
investigated in the last years. A lot of impressive progresses
have been made in this field of research based mainly on the
integrability structures discovered on both sides of the
correspondence. The most studied example is the correspondence
between type IIB string theory on ${\rm AdS}_5\times S^5$ target
space and the ${\cal N}=4$ super Yang-Mills theory (SYM) in four
space-time dimensions. However, many other cases are also of
interest, and have been investigated intensively (for recent
review on the AdS/CFT duality, see \cite{RO}).

Different classical string solutions play important role in
checking and understanding the AdS/CFT correspondence
\cite{AAT1012}. To establish relations with the dual gauge theory,
one has to take the semiclassical limit of {\it large} conserved
charges \cite{GKP02}. An important example of such string solution is
the so called "giant magnon" living in the $R_t\times S^2$ subspace of ${\rm AdS}_5\times S^5$,
discovered by Hofman and Maldacena \cite{HM06}.
It gave a strong support for the conjectured all-loop
$SU(2)$ spin chain, arising in the dual $\mathcal{N}=4$ SYM, and
made it possible to get a deeper insight in the AdS/CFT duality.
Characteristic feature of this solution is that the string energy
$E$ and the angular momentum $J_1$ go to infinity, but the
difference $E-J_1$ remains finite and it is related to the momentum of
the magnon excitations in the dual spin chain in $\mathcal{N}=4$
SYM. This string configuration have been extended to the case of
{\it dyonic} giant magnon, being solution for a string moving on
$R_t\times S^3$ and having second nonzero angular momentum $J_2$
\cite{ND06}. Further extension to $R_t\times S^5$ have been also
worked out in \cite{KRT06}. It was also shown there that such type
of string solutions can be obtained by reduction of the string
dynamics to the Neumann-Rosochatius integrable system, by using a
specific ansatz.

An interesting issue to solve is to find the {\it finite-size
effect}, i.e. $J_1$ large, but finite, related to the wrapping
interactions in the dual field theory \cite{Janikii}. For (dyonic)
giant magnons living in ${\rm AdS}_5\times S^5$ this was done in
\cite{AFZ06,HS0801}. The corresponding string solutions, along
with the (leading) finite-size corrections to their dispersion
relations have been found.

Another issue is to go beyond the spectral problem by computing different correlation functions.
For two-, three- and four-point correlators a lot of interesting results have been obtained
\footnote{The recent activity in computing the semiclassical correlation functions
in the framework of the AdS/CFT duality was initiated in
\cite{JSW1002}-\cite{rt10}.}.
Further investigations on the problem include the finite-size effects on some of them\cite{AB1105}-\cite{B1304}.

Two years ago a new {\it integrable} deformation
($\eta$-deformation) of the $AdS_5 \times S^5$ superstring action
has been discovered \cite{DMV0913}\footnote{For the recent
investigations related to this integrable deformation of $AdS_5$ x
$S^5$ see \cite{ABF1312}-\cite{HT1504}.}. The bosonic part of the
superstring sigma model Lagrangian on this $\eta$-deformed
background was determined in \cite{ABF1312}. From it one can
extract the background metric and $B$-field \cite{AP2014}.

The spectrum of a string moving on $\eta$-deformed $AdS_5 \times
S^5$ was considered in \cite{ALT1403}. This is done by treating
the corresponding worldsheet theory as integrable field theory. In
particular, it was found that the dispersion relation for the
infinite-size giant magnons \cite{HM06} on this background, in the
large string tension limit when $g \to \infty$ is given by
\bea\label{14} E=\frac{2g\sqrt{1+\tilde{\eta}^2}}{\tilde{\eta}}
\mbox{arcsinh}\left(\tilde{\eta} \sin\frac{p}{2}\right),\eea where
$\tilde{\eta}$ is related to the deformation parameter $\eta$
according to\footnote{We changed the notation $\kappa$ in
\cite{ALT1403} to $\tilde{\eta}$ because we use $\kappa$ for other
purposes.} \bea\label{ek} \tilde{\eta}=\frac{2
\eta}{1-\eta^2}.\eea

The result (\ref{14}) has been extended in \cite{AP2014} to the
case of finite-size giant magnons. The corresponding dispersion
relation is given by \bea\label{fr} E_s-J_1= 2 g
\sqrt{1+\tilde{\eta}^2} \left[\frac{1}{\tilde{\eta}}
\mbox{arcsinh}\left(\tilde{\eta}
\sin\frac{p}{2}\right)-\frac{(1+\tilde{\eta}^2)
\sin^3\frac{p}{2}}{4 \sqrt{1+\tilde{\eta}^2 \sin^2\frac{p}{2}}}\
\epsilon\right],\eea where
\bea\label{eps} \epsilon =16\
\exp\left[-\left(\frac{J_1}{g}
+\frac{2\sqrt{1+\tilde{\eta}^2}}{\tilde{\eta}}\mbox{arcsinh}\left(\tilde{\eta}
\sin\frac{p}{2}\right)
\right)\sqrt{\frac{1+\tilde{\eta}^2\sin^2\frac{p}{2}}{\left(1+\tilde{\eta}^2\right)\sin^2\frac{p}{2}}}
\right].\eea

Here we are going to obtain, from the string theory viewpoint,
some semiclassical three-point correlation functions in the
$\eta$-deformed $AdS_5\times S^5$.

The paper is organized as follows. In Sec. 2 we describe the
finite-size giant magnon solution on the $\eta$-deformed
$AdS_5\times S^5$. In Sec. 3, we derive the exact semiclassical
structure constants for the case when the ``heavy'' string states
are {\it finite-size} giant magnons, carrying one angular
momentum, for three different choices of the ``light'' state:
primary scalar operators, dilaton operator with nonzero momentum,
singlet scalar operators on higher string levels. Sec. 4 is
devoted to our concluding remarks.

\setcounter{equation}{0}
%%%%%%%%%%%%%5%%%%%%%%%%%%%%%%%%%%%%%%%%%%%%%%%%%%%%%%%%%%%%%%%%%%%%%%%%%%%%%%
\section{Giant magnons on $\eta$-deformed $AdS_5\times S^5$}
%%%%%%%%%%%%%5%%%%%%%%%%%%%%%%%%%%%%%%%%%%%%%%%%%%%%%%%%%%%%%%%%%%%%%%%%%%%%%%

Giant magnons live in the $R_t\times S^2_{\eta}$ subspace of the
$\eta$-deformed $AdS_5\times S^5$. The background seen by the
string moving in the $R_t\times S^2_\eta$ subspace can be written
as \cite{AP2014} \bea\nn &&g_{tt}=-1,\h
g_{\phi_1\phi_1}=\sin^2\theta,
\\ \label{fb} &&g_{\theta\theta}=\frac{1}{1+\tilde{\eta}^2 \sin^2\theta}.\eea

By using a specific anzatz for the string embedding \cite{KRT06},
one can find the following string solution
\cite{AP2014}\footnote{See \cite{AR1406} where it was shown that
the bosonic spinning strings on the $\eta$-deformed $AdS_5 x S^5$
background are naturally described as periodic solutions of a
novel finite-dimensional integrable system which can be viewed as
a deformation of the Neumann model.} \bea\label{chis} \chi(\xi)=
\frac{\chi_\eta\chi_p\ \mathbf{dn}^2(x,m)}{\chi_p \
\mathbf{dn}^2(x,m)+\chi_\eta-\chi_p},\eea

\bea\label{f1s} &&\phi_1(\tau,\sigma)= \omega_1
\tau+\frac{1}{\tilde{\eta}\alpha\omega_1^2(\chi_\eta-1)
\sqrt{(\chi_\eta-\chi_m)\chi_p}} \times
\\ \nn &&\Bigg\{\Big[\beta\left(\kappa^2+\omega_1^2(\chi_\eta-1)\right)
\Big]\
\mathbf{F}\left(\arcsin\sqrt{\frac{(\chi_\eta-\chi_m)(\chi_p-\chi)}{(\chi_p-\chi_m)(\chi_\eta-\chi)}},
m\right)
\\ \nn &&-\frac{(\chi_\eta-\chi_p)\beta\kappa^2}{1-\chi_p}\
\mathbf{\Pi}\left(\arcsin\sqrt{\frac{(\chi_\eta-\chi_m)(\chi_p-\chi)}{(\chi_p-\chi_m)(\chi_\eta-\chi)}}
,-\frac{(\chi_\eta-1)(\chi_p-\chi_m)}{(1-\chi_p)(\chi_\eta-\chi_m)},m\right)\Bigg\}.\eea

Here $\chi=\cos^2\theta$, where $\theta$ is the non-isometric
angle on the deformed sphere $S^2_{\eta}$, while $\phi_1$ is the
isometric angle on it. $\mathbf{dn}(x,m)$ is one of the Jacobi
elliptic functions, $\mathbf{F}$ and $\mathbf{\Pi}$ are the
incomplete elliptic integrals of first and third kind. \bea\nn
&&x=\frac{\tilde{\eta} \alpha
\omega_1\sqrt{(\chi_\eta-\chi_m)\chi_p}} {\alpha^2-\beta^2}\ \xi ,
\\ \nn &&m= \frac{(\chi_p-\chi_m)\chi_\eta}{(\chi_\eta-\chi_m)\chi_p},\eea
and $\chi_\eta >\chi_p>\chi_m$ are the roots of the equation $d
\chi/d \xi=0$.

Now, let us present the expressions for the conserved charges (the
string energy $E_s$ and the angular momentum $J_1$) and also the
worldsheet momentum $p$ equal to the angular difference
$\Delta\phi_1$, since we are going to use it \cite{AP2014}:
\bea\label{Esi}
E_s=\frac{T}{\tilde{\eta}}\left(1-\frac{\beta^2}{\alpha^2}\right)
\frac{\kappa}{\omega_1} \int_{\chi_m}^{\chi_p}
\frac{d\chi}{\sqrt{(\chi_\eta-\chi)(\chi_p-\chi)(\chi-\chi_m)\chi}},\eea
\bea\label{J1i} J_1&=&\frac{T}{\tilde{\eta}}
\left[\left(1-\frac{\beta^2\kappa^2}{\alpha^2\omega_1^2}\right)
\int_{\chi_m}^{\chi_p}
\frac{d\chi}{\sqrt{(\chi_\eta-\chi)(\chi_p-\chi)(\chi-\chi_m)\chi}}\right.
\\ \nn &&-\left.\int_{\chi_m}^{\chi_p}
\frac{\chi
d\chi}{\sqrt{(\chi_\eta-\chi)(\chi_p-\chi)(\chi-\chi_m)\chi}}\right],\eea

\bea\label{adi} \Delta\phi_1 &=&\frac{1}{\tilde{\eta}}
\Bigg[\frac{\beta}{\alpha}\int_{\chi_m}^{\chi_p}
\frac{d\chi}{\sqrt{(\chi_\eta-\chi)(\chi_p-\chi)(\chi-\chi_m)\chi}}
\\ \nn &&-\frac{\beta\kappa^2}{\alpha\omega_1^2}\int_{\chi_m}^{\chi_p}
\frac{d\chi}{(1-\chi)\sqrt{(\chi_\eta-\chi)(\chi_p-\chi)(\chi-\chi_m)\chi}}\Bigg].\eea
Solving the integrals in (\ref{Esi})-(\ref{adi}) and introducing
the notations \bea\label{not} v=-\frac{\beta}{\alpha},\
W=\frac{\kappa^2}{\omega_1^2},\
\epsilon=\frac{(\chi_\eta-\chi_p)\chi_m}{(\chi_\eta-\chi_m)\chi_p},\eea
we finally obtain \bea\label{Esf} E_s=\frac{2T}{\tilde{\eta}}
\frac{(1-v^2)\sqrt{W}}{\sqrt{(\chi_\eta-\chi_m)\chi_p}} \
\mathbf{K}(1-\epsilon),\eea \bea\label{J1f}
J_1&=&\frac{2T}{\tilde{\eta}\sqrt{(\chi_\eta-\chi_m)\chi_p}}
\Bigg[\left(1-v^2W-\chi_\eta\right)\ \mathbf{K}(1-\epsilon)
\\ \nn &&+(\chi_\eta-\chi_p)\ \mathbf{\Pi}\left(\frac{\chi_p-\chi_m}{\chi_\eta-\chi_m},
1-\epsilon\right)\Bigg],\eea

\bea\label{adf} \Delta\phi_1 &=&
\frac{2}{\tilde{\eta}\sqrt{(\chi_\eta-\chi_m)\chi_p}}\times
\\ \nn &&\Bigg\{\frac{v W}{(\chi_\eta-1)(1-\chi_p)}\Bigg[(\chi_\eta-\chi_p)\
\mathbf{\Pi}\left(-\frac{(\chi_\eta-1)(\chi_p-\chi_m)}{(\chi_\eta-\chi_m)(1-\chi_p)},
1-\epsilon\right)
\\ \nn &&-(1-\chi_p)\ \mathbf{K}(1-\epsilon)\Bigg]-v \ \mathbf{K}(1-\epsilon)\Bigg\},\eea
where $\mathbf{K}$ and $\mathbf{\Pi}$ are the complete elliptic
integrals of first and third kind.

\setcounter{equation}{0}
%%%%%%%%%%%%%5%%%%%%%%%%%%%%%%%%%%%%%%%%%%%%%%%%%%%%%%%%%%%%%%%%%%%%%%%%%%%%%%
\section{Semiclassical three-point correlation functions}
%%%%%%%%%%%%%5%%%%%%%%%%%%%%%%%%%%%%%%%%%%%%%%%%%%%%%%%%%%%%%%%%%%%%%%%%%%%%%%
It is known that the correlation functions of any conformal field
theory can be determined  in principle in terms of the basic
conformal data $\{\Delta_i,C_{ijk}\}$, where $\Delta_i$ are the
conformal dimensions defined by the two-point correlation
functions
\begin{equation}\nn
\left\langle{\cal O}^{\dagger}_i(x_1){\cal O}_j(x_2)\right\rangle=
\frac{C_{12}\delta_{ij}}{|x_1-x_2|^{2\Delta_i}}
\end{equation}
and $C_{ijk}$ are the structure constants in the operator product
expansion
\begin{equation}\nn
\left\langle{\cal O}_i(x_1){\cal O}_j(x_2){\cal
O}_k(x_3)\right\rangle=
\frac{C_{ijk}}{|x_1-x_2|^{\Delta_1+\Delta_2-\Delta_3}
|x_1-x_3|^{\Delta_1+\Delta_3-\Delta_2}|x_2-x_3|^{\Delta_2+\Delta_3-\Delta_1}}.
\end{equation}
Therefore, the determination of the initial conformal data for a
given conformal field theory is the most important step in the
conformal bootstrap approach.

The three-point functions of two ``heavy'' operators and a ``light''
operator can be approximated by a supergravity vertex operator
evaluated at the ``heavy'' classical string configuration
\cite{rt10,Hernandez2}: \bea \nn \langle
V_{H}(x_1)V_{H}(x_2)V_{L}(x_3)\rangle=V_L(x_3)_{\rm classical}.
\eea For $\vert x_1\vert=\vert x_2\vert=1$, $x_3=0$, the
correlation function reduces to \bea \nn \langle
V_{H}(x_1)V_{H}(x_2)V_{L}(0)\rangle=\frac{C_{123}}{\vert
x_1-x_2\vert^{2\Delta_{H}}}. \eea Then, the normalized structure
constants \bea \nn \mathcal{C}=\frac{C_{123}}{C_{12}} \eea can be
found from \bea \label{nsc} \mathcal{C}=c_{\Delta}V_L(0)_{\rm
classical}, \eea were $c_{\Delta}$ is the normalized constant of
the corresponding ``light'' vertex operator.
Actually, we are going to compute the normalized structure constants (\ref{nsc}).

Let us mention that in the case of $\eta$-deformed $AdS_5\times
S^5$ it is not clear from the beginning when the approach
described above for computing the semiclassical three-point
correlation functions from the string theory can be applied. There
are several reasons for this.

First of all, the complete supergravity solution and its
symmetries are not known for the case of $\eta$-deformed
$AdS_5\times S^5$ \footnote{It was proven in \cite{LRT1411} that
in the $\eta$-deformed $AdS_2 x S^2$ subspace the deformed metric
can be extended to a full supergravity solution.}. Therefore, one
can not make a proposal for a dual gauge theory. Since the $AdS_5$
subspace is also deformed, the corresponding gauge theory is not
conformal. As it was pointed out in \cite{ABF1312},  one may
expect that this field theory is a non-commutative deformation of
${\cal N}=4$ SYM.

However, the case of giant magnons is very special. Since these
are strings moving in the $R_t\times S^2_{\eta}$ subspace and
there is no deformation along the time direction $t$ in $AdS$, we
suppose that we can use the above formulas, based on the existence
of conformal symmetry, for the computation of semiclassical
three-point correlation functions in the $\eta$-deformed case. The
effect of the $\eta$-deformation on the sphere can be taken into
account in the same way as for the case of the $TsT$-deformation
of $AdS_5\times S^5$ \cite{F2005}, as was done in
\cite{AB1106}-\cite{B1108}.

%%%%%%%%%%%%%5%%%%%%%%%%%%%%%%%%%%%%%%%%%%%%%%%%%%%%%%%%%%%%%%%%%%%%%%%%%%%%%%
\subsection{String vertices}
%%%%%%%%%%%%%5%%%%%%%%%%%%%%%%%%%%%%%%%%%%%%%%%%%%%%%%%%%%%%%%%%%%%%%%%%%%%%%%
Let us first recall the situation in the undeformed $AdS_5\times
S^5$ case.

We denote with $Y$, $X$ the coordinates in $AdS_5$ and $S^5$ parts
of the background $AdS_5\times S^5$: \bea\nn &&Y_1+iY_2=\sinh\rho\
\sin\eta\ e^{i\varphi_1},\\ \nn &&Y_3+iY_4=\sinh\rho\ \cos\eta\
e^{i\varphi_2},\\ \nn &&Y_5+iY_0=\cosh\rho\ e^{it}. \eea The
coordinates $Y$ are related to the Poincare coordinates by \bea
\nn &&Y_m=\frac{x_m}{z},\\ \nn
&&Y_4=\frac{1}{2z}\left(x^mx_m+z^2-1\right), \\ \nn
&&Y_5=\frac{1}{2z}\left(x^mx_m+z^2+1\right), \eea where $x^m
x_m=-x_0^2+x_ix_i$, with $m=0,1,2,3$ and $i=1,2,3$. We
parameterize $S^5$ as in \cite{Hernandez2}.

Euclidean continuation of the time-like directions to $t_e = it$ ,
$Y_{0e} = iY_0$ , $x_{0e} = ix_0$, will allow the classical
trajectories to approach the $AdS_5$ boundary $z=0$ when $\tau_e
\rightarrow \pm\infty$, and to compute the corresponding
correlation functions.

For our purposes, we need to know the string vertices for the
following ``light'' states:
\begin{enumerate}
\item{Primary scalar operators: $V_L=V^{pr}_j$}
\item{Dilaton
operator: $V_L=V^d_j$}
\item{Singlet scalar operators on higher
string levels: $V_L= V^q$}
\end{enumerate}

According to \cite{rt10}, these (unintegrated) vertices are given
by \bea \label{prv} V^{pr}_j&=&\left(Y_4+Y_5\right)^{-\Delta_{pr}}
\left(X_1+iX_2\right)^j
\\ \nn
&&\left[z^{-2}\left(\p x_{m}\bar{\p}x^{m}-\p z\bar{\p}z\right) -\p
X_{k}\bar{\p}X_{k}\right],\h \Delta_{pr}=j,\eea
\bea \label{dv}
V^d_j&=&\left(Y_4+Y_5\right)^{-\Delta_d} \left(X_1+iX_2\right)^j
\\ \nn
&&\left[z^{-2}\left(\p x_{m}\bar{\p}x^{m}+\p z\bar{\p}z\right) +\p
X_{k}\bar{\p}X_{k}\right],\h \Delta_d=4+j,\eea
\bea\label{Vq}
V^q=(Y_4+Y_5)^{- \Delta_q} (\p X_k \bar{\p} X_k)^q ,\h
\Delta_q=2\left(\sqrt{(q-1)\sqrt{\lambda}+1-\frac{1}{2}q(q-1)}+1\right).\eea

The solution for giant magnons with infinite or finite-size in the
Euclidean $AdS$ can be written as ($t=\kappa\tau$,\
$i\tau=\tau_e$) \bea\label{adssol} &&x_{0e}=\tanh(\kappa\tau_e),\h
x_i=0,\h z=\frac{1}{\cosh(\kappa\tau_e)}.\eea

Let us point out that for the $\eta$-deformed case the solution is
given again by (\ref{adssol}). Therefore, the contribution to the
above three vertices from $AdS$ is the same as in the undeformed
case.

In order to take into account the $\eta$-deformations on the above
vertices for giant magnons, we propose to deform them in an
appropriate way. As one can see, all of them contain the term $\p
X_{k}\bar{\p}X_{k}$. Since for the undeformed case it is
proportional to the string Lagrangian on $S^2$ computed on the
corresponding string solution, it is natural to suppose that for
the $\eta$-deformed case its contribution should be computed on
$S^2_\eta$. Namely this approach we will use further on.

%%%%%%%%%%%%%5%%%%%%%%%%%%%%%%%%%%%%%%%%%%%%%%%%%%%%%%%%%%%%%%%%%%%%%%%%%%%%%%
\subsection{ Finite-size giant magnons and primary scalar operators}
%%%%%%%%%%%%%5%%%%%%%%%%%%%%%%%%%%%%%%%%%%%%%%%%%%%%%%%%%%%%%%%%%%%%%%%%%%%%%%
According to \cite{B1107},  the normalized structure constants for
the undeformed case can be written as \bea\label{c3pr}
\mathcal{C}^{pr,j}&=&c_{\Delta}^{pr,j}
\left[\int_{-\infty}^{\infty}d\tau_e
\frac{\kappa^2}{\cosh^{j}(\kappa\tau_e)}
\left(\frac{2}{\cosh^2(\kappa\tau_e)}-1\right)
\int_{-L}^{L}d\sigma\chi^{\frac{j}{2}} \right.
\\ \nn &&-\left.\int_{-\infty}^{\infty} \frac{d\tau_e}{\cosh^{j}(\kappa\tau_e)}
\int_{-L}^{L}d\sigma\chi^{\frac{j}{2}}\p X_K \bar{\p} X_K\right],
\h j\geq 1\eea where $t=\kappa \tau_e$ is the Euclidean $AdS$
time, the term $\p X_K \bar{\p} X_K$ is proportional to the string
Lagrangian on $S^2$ computed on the finite-size giant magnon
solution living in the $R_t\times S^2$ subspace, and $\chi=\cos^2
\theta$ ($\theta$ is the non-isometric angle on $S^2$). The
parameter $L$ gives the size of the giant magnon.

For giant magnon solution on the $\eta$-deformed background the
contribution from the $AdS$ subspace is the same. So, the
integration over $\tau_e$ leads to \bea\label{c3pri}
\mathcal{C}_{\eta}^{pr,j}&=&c_{\Delta}^{pr,j}\sqrt{\pi}
\frac{\Gamma (\frac{j}{2})}{\Gamma (\frac{1+j}{2})}
 \left[\frac{j-1}{j+1}\kappa \int_{-L}^{L}d\sigma\chi^{\frac{j}{2}}-\frac{1}{\kappa}
\int_{-L}^{L}d\sigma\chi^{\frac{j}{2}}\p X_K \bar{\p} X_K
\right].\eea

To take into account the $\eta$-deformation of the two-sphere, we
should compute $\p X_K \bar{\p} X_K$ on $S^2_\eta$. By using some
of the results obtained in \cite{AP2014}, one can show that
\bea\label{ecs} \p X_K \bar{\p} X_K =
-\frac{2-(1+v^2)\kappa^2-2\chi}{1-v^2},\eea where $v$ is the
worldsheet velocity.

The computations are given in an Appendix. Here we present the
final result \bea\label{c3prsf}
\mathcal{C}_{\eta}^{pr,j}&=&\frac{2\pi^{\frac{3}{2}}c_{\Delta}^{pr,j}\Gamma
(\frac{j}{2}) (1-v^2 \kappa^2)^{\frac{j-1}{2}}} {\Gamma
(\frac{1+j}{2})\sqrt{\kappa^2(1+\tilde{\eta}^2 \kappa^2)}}
\Bigg\{\left[1-\frac{(1+jv^2)\kappa^{2}}{1+j}\right]\times
\\ \nn &&F_1 \left(\frac{1}{2},\frac{j}{2},\frac{1-j}{2};1;\frac{\tilde{\eta}^2(1-v^2)\kappa^2}
{1+\tilde{\eta}^2\kappa^2},\frac{(1+\tilde{\eta}^2)(1-v^2)\kappa^2}
{(1+\tilde{\eta}^2\kappa^2)(1-v^2\kappa^2)}\right)
\\ \nn &&-(1-v^2\kappa^2) F_1 \left(\frac{1}{2},\frac{2+j}{2},-\frac{1+j}{2};1;\frac{\tilde{\eta}^2(1-v^2)\kappa^2}
{1+\tilde{\eta}^2\kappa^2},\frac{(1+\tilde{\eta}^2)(1-v^2)\kappa^2}
{(1+\tilde{\eta}^2\kappa^2)(1-v^2\kappa^2)}\right)\Bigg\} ,\eea
where $F_1$ is one of the hypergeometric functions of two
variables ($Appell F_1$).

Next, we would like to compare (\ref{c3prsf}) with the known
expression for the undeformed case \cite{B1107}. To this end, we
take the limit $\tilde{\eta}\to 0$ and by using that \cite{PBMv3}
\bea\nn F_1 \left(a,b_1,b_2;c;0,z_2\right)=\
{}_2F_1(a,b_2;c;z_2),\eea where ${}_2F_1(a,b_2;c;z_2)$ is the
Gauss' hypergeometric function, we find \bea\nn
\mathcal{C}^{pr,j}&=&
\frac{2\pi^{\frac{3}{2}}c_{\Delta}^{pr,j}\Gamma\left(2+\frac{j}{2}\right)(1-v^2
\kappa^2)^{\frac{j-1}{2}}} {\kappa
\Gamma\left(\frac{5+j}{2}\right)} \Bigg[(1-v^2
\kappa^2){}_2F_1\left(\frac{1}{2},-\frac{1+j}{2};1;,\frac{(1-v^2)\kappa^2}
{1-v^2\kappa^2}\right)
\\ \nn &&-(1-\kappa^2){}_2F_1\left(\frac{1}{2},\frac{1-j}{2};1;\frac{(1-v^2)\kappa^2)}{1-v^2 \kappa^2}\right)\Bigg].\eea
This is exactly the same result found in \cite{B1107} for $u=0$
(finite-size giant magnons with one nonzero angular momentum) as
it should be \footnote{In our notations $W=\kappa^2$ and
(\ref{pars}) is taken into account.}.

Let us also give an example for the simplest case when $j=1$. In
that case (\ref{c3prsf}) reduces to
\bea\nn\mathcal{C}_{\eta}^{pr,1} &=&\frac{2 \pi c_{\Delta}^{pr,1}}
{\tilde{\eta}^2\sqrt{\kappa^2(1+\tilde{\eta}^2\kappa^2)}}
\Bigg[2(1+\tilde{\eta}^2\kappa^2)\mathbf{E}\left(\tilde{\eta}^2
\frac{(1-v^2)\kappa^2}{1+\tilde{\eta}^2\kappa^2}\right)
\\ \nn &&-(2+(1+v^2)\tilde{\eta}^2\kappa^2)\mathbf{K}\left(\tilde{\eta}^2
\frac{(1-v^2)\kappa^2}{1+\tilde{\eta}^2\kappa^2}\right)\Bigg],\eea
where $\mathbf{K}$ and $\mathbf{E}$ are the complete elliptic
integrals of first and second kind. In the limit $\tilde{\eta}\to
0$, $\mathcal{C}_{\eta}^{pr,1}\to 0$. The same result can be
obtained from (2.13) in \cite{B1107} by taking $u\to 0, j\to 1$.

%%%%%%%%%%%%%5%%%%%%%%%%%%%%%%%%%%%%%%%%%%%%%%%%%%%%%%%%%%%%%%%%%%%%%%%%%%%%%%
\subsection{ Finite-size giant magnons and dilaton with nonzero momentum}
%%%%%%%%%%%%%5%%%%%%%%%%%%%%%%%%%%%%%%%%%%%%%%%%%%%%%%%%%%%%%%%%%%%%%%%%%%%%%%
The case of finite-size giant magnons and dilaton with zero momentum $(j=0)$ have been considered in \cite{AB1412}.
Here we will be interested in the case when $j>0$.

According to \cite{B1107} the semiclassical normalized structure constants for the undeformed case are given by
\bea\nn
\mathcal{C}^{d,j}&=&c_\Delta^{d,j}\int_{-\infty}^{\infty}\frac{d\tau_e}{\cosh^{4+j}(\kappa \tau_e)}
\int_{-L}^{L}d \sigma \chi^{\frac{j}{2}}\left(\kappa^{2}+\p X_K \bar{\p} X_K\right)
\\ \nn &&=c_\Delta^{d,j}\frac{\sqrt{\pi}\Gamma(2+\frac{j}{2})}{\kappa\Gamma(\frac{5+j}{2})}
\int_{-L}^{L}d \sigma \chi^{\frac{j}{2}}\left(\kappa^{2}+\p X_K \bar{\p} X_K\right).\eea

In order to take into account the $\eta$-deformation, we must use
(\ref{ecs}), (\ref{cp}), (\ref{pars}). Working in the same way as
in the previous subsection (see the Appendix), one obtains the
following result for $\mathcal{C}_{\eta}^{d,j}$: \bea\label{etadj}
\mathcal{C}_{\eta}^{d,j}&=&\frac{\pi^{\frac{1}{2}}c_\Delta^{d,j}
\Gamma\left(2+\frac{j}{2}\right)(1-v^2)}
{\Gamma\left(\frac{5+j}{2}\right)\tilde{\eta} \kappa}
\int_{\chi_m}^{\chi_p}d\chi
\frac{\chi^{\frac{j-1}{2}}\left[\kappa^2-\frac{2-(1+v^2)\kappa^2-2\chi}{1-v^2}\right]}
{\sqrt{(\chi_\eta-\chi)(\chi_p-\chi)(\chi-\chi_m)\chi}}
\\ \nn &=&2\pi^{\frac{3}{2}}c_\Delta^{d,j}\frac{\Gamma\left(2+\frac{j}{2}\right)}{\Gamma\left(\frac{5+j}{2}\right)}
\frac{\chi_m^{\frac{j+1}{2}}}{\tilde{\eta}\sqrt{(\chi_\eta-\chi_m)(1-\chi_m)}}\times
\\ \nn && \Bigg[F_1 \left(\frac{1}{2},\frac{1}{2},-\frac{j+1}{2};1;\frac{\chi_p-\chi_m}
{\chi_\eta-\chi_m},-\frac{\chi_p-\chi_m}{\chi_m}\right)
\\ \nn &&-
F_1 \left(\frac{1}{2},\frac{1}{2},-\frac{j-1}{2};1;\frac{\chi_p-\chi_m}
{\chi_\eta-\chi_m},-\frac{\chi_p-\chi_m}{\chi_m}\right)\Bigg]
\\ \nn
&=&\frac{2 \pi^{\frac{3}{2}}c_\Delta^{d,j}
\Gamma\left(2+\frac{j}{2}\right)(1-v^2 \kappa^2)^{\frac{j-1}{2}}}
{\Gamma\left(\frac{5+j}{2}\right)\sqrt{\kappa^2(1+\tilde{\eta}^2 \kappa^2)}}\times
\\ \nn
&&\Bigg[(1-v^2 \kappa^2)F_1 \left(\frac{1}{2},\frac{2+j}{2},-\frac{1+j}{2};1;\frac{\tilde{\eta}^2(1-v^2)\kappa^2}
{1+\tilde{\eta}^2\kappa^2},\frac{(1+\tilde{\eta}^2)(1-v^2)\kappa^2}
{(1+\tilde{\eta}^2\kappa^2)(1-v^2\kappa^2)}\right)
\\ \nn &&-(1-\kappa^2)F_1 \left(\frac{1}{2},\frac{j}{2},\frac{1-j}{2};1;\frac{\tilde{\eta}^2(1-v^2)\kappa^2}
{1+\tilde{\eta}^2\kappa^2},\frac{(1+\tilde{\eta}^2)(1-v^2)\kappa^2}
{(1+\tilde{\eta}^2\kappa^2)(1-v^2\kappa^2)}\right)\Bigg].\eea

Now we take the limit $\tilde{\eta}\to 0$ in (\ref{etadj}) and obtain
\bea\label{dud}
\mathcal{C}^{d,j}&=&\frac{2 \pi^{\frac{3}{2}}c_\Delta^{d,j}
\Gamma\left(2+\frac{j}{2}\right)(1-v^2 \kappa^{2})^{\frac{j-1}{2}}}{\kappa
\Gamma\left(\frac{5+j}{2}\right)} \times
\\ \nn
&&\Bigg[(1-v^2 \kappa^2){}_2F_1\left(\frac{1}{2},-\frac{1+j}{2};1;\frac{(1-v^2)\kappa^2)}{1-v^2 \kappa^2}\right)
\\ \nn
&&-(1-\kappa^2){}_2F_1\left(\frac{1}{2},\frac{1-j}{2};1;\frac{(1-v^2)\kappa^2)}{1-v^2 \kappa^2}\right)\Bigg] .\eea
This is exactly what was found in \cite{B1107} for $u=0$, as it should be.

Let us also say that in the particular case when $j=1$, (\ref{etadj}) simplifies to
\bea\nn \mathcal{C}_{\eta}^{d,1}=\frac{3\pi c_\Delta^{d,1}\sqrt{\kappa^2(1+\tilde{\eta}^2 \kappa^2)}}
{2 \tilde{\eta}^2 \kappa^2}\Bigg[\mathbf{K}\left(\tilde{\eta}^2
\frac{(1-v^2)\kappa^2}{1+\tilde{\eta}^2\kappa^2}\right)-\mathbf{E}\left(\tilde{\eta}^2
\frac{(1-v^2)\kappa^2}{1+\tilde{\eta}^2\kappa^2}\right)\Bigg] .\eea
In the limit $\tilde{\eta}\to 0$, $\mathcal{C}_{\eta}^{d,1}$ becomes
\bea\nn\mathcal{C}^{d,1}=\frac{3}{8}\pi^2  c_\Delta^{d,1}\kappa (1-v^2).\eea

%%%%%%%%%%%%%5%%%%%%%%%%%%%%%%%%%%%%%%%%%%%%%%%%%%%%%%%%%%%%%%%%%%%%%%%%%%%%%%
\subsection{ Finite-size giant magnons and singlet scalar operators on higher string levels}
%%%%%%%%%%%%%5%%%%%%%%%%%%%%%%%%%%%%%%%%%%%%%%%%%%%%%%%%%%%%%%%%%%%%%%%%%%%%%%
According to \cite{B1108} the normalized structure constant for the undeformed case is given by
\bea\label{Cq} \mathcal{C}^q&=&c_{\Delta}^{q}\int_{-\infty}^{\infty}\frac{d\tau_e}{\cosh^{\Delta_q}(\kappa \tau_e)}
\int_{-L}^{L}d \sigma \left(\p X_K \bar{\p} X_K\right)^{q}
\\ \nn &=&c_{\Delta}^{q}\frac{\sqrt{\pi}}{\kappa} \frac{\Gamma\left(\frac{\Delta_q}{2}\right)}
{\Gamma\left(\frac{\Delta_q+1}{2}\right)}\int_{-L}^{L}d \sigma \left(\p X_K \bar{\p} X_K\right)^{q}.\eea
Here the parameter $q$ is related to the string level $n$ as $q=n+1\geq 1$ and
\bea\label{dq1} \Delta_q=2\left(1+\sqrt{(q-1)\sqrt{\lambda}+1-\frac{1}{2}q(q-1)}\right),\eea
where $\lambda$ is the 't Hooft coupling constant in the dual gauge theory.
Taking into account that the string tension $T$ is related to the 't Hooft coupling as
\bea\nn T=\frac{\sqrt{\lambda}}{2\pi},\eea
(\ref{dq1}) can be rewritten as
\bea\label{dq} \Delta_q=2\left(1+\sqrt{2\pi T(q-1)+1-\frac{1}{2}q(q-1)}\right).\eea

For the $\eta$-deformed case we have \cite{ABF1312} \bea\nn
T=g\sqrt{1+\tilde{\eta}^2}.\eea So, \bea\label{dqe}
\Delta_q^\eta=2\left(1+\sqrt{2\pi
g\sqrt{1+\tilde{\eta}^2}(q-1)+1-\frac{1}{2}q(q-1)}\right).\eea In
addition, to compute the integral over $\sigma$ in (\ref{Cq}), we
have to use (\ref{ecs}), (\ref{cp}), (\ref{pars}). Thus
\bea\label{qf} \mathcal{C}^q_\eta &=&
c_{\Delta}^{q}\frac{\sqrt{\pi}}{\kappa}
\frac{\Gamma\left(\frac{\Delta_q^\eta}{2}\right)}
{\Gamma\left(\frac{\Delta_q^\eta+1}{2}\right)}\frac{(-1)^q}{\tilde{\eta}(1-v^2)^{q-1}}
\int_{\chi_m}^{\chi_p}d \chi
\frac{\left[2-(1+v^2)\kappa^2-2\chi\right]^{q}}
{\sqrt{(\chi_\eta-\chi)(\chi_p-\chi)(\chi-\chi_m)}\chi}
\\ \nn &=& c_{\Delta}^{q}\frac{\pi^{\frac{3}{2}}}{\kappa} \frac{\Gamma\left(\frac{\Delta_q^\eta}{2}\right)}
{\Gamma\left(\frac{\Delta_q^\eta+1}{2}\right)}\frac{(-1)^q\left[2-(1+v^2)\kappa^2\right]^{q}}
{\tilde{\eta}(1-v^2)^{q-1}\sqrt{\chi_\eta-\chi_m}}\times
\\ \nn && \sum_{k=0}^{q}\frac{q!}{k!(q-k)!}\left[-\frac{1}{1-\frac{1}{2}(1+v^2)\kappa^2}\right]^k \chi_m^{k-\frac{1}{2}}
F_1 \left(\frac{1}{2},\frac{1}{2},\frac{1}{2}-k;1;\frac{\chi_p-\chi_m}{\chi_\eta-\chi_m},
-\frac{\chi_p-\chi_m}{\chi_m}\right)
\\ \nn &=&c_{\Delta}^{q}\pi^{\frac{3}{2}}\frac{\Gamma\left(\frac{\Delta_q^\eta}{2}\right)}
{\Gamma\left(\frac{\Delta_q^\eta+1}{2}\right)}\frac{(-1)^q\left[2-(1+v^2)\kappa^2\right]^{q}}
{(1-v^2)^{q-1}\sqrt{\kappa^2(1+\tilde{\eta}^2 \kappa^2)(1-v^2 \kappa^2)}}\sum_{k=0}^{q}\frac{q!}{k!(q-k)!}
\\ \nn &&  \times \left[-\frac{1-v^2\kappa^2}{1-\frac{1}{2}(1+v^2)\kappa^2}\right]^k
F_1 \left(\frac{1}{2},k,\frac{1}{2}-k;1;\frac{\tilde{\eta}^2(1-v^2)\kappa^2}
{1+\tilde{\eta}^2\kappa^2},\frac{(1+\tilde{\eta}^2)(1-v^2)\kappa^2}
{(1+\tilde{\eta}^2\kappa^2)(1-v^2\kappa^2)}\right).\eea

In order to compare with the undeformed case, we take the limit $\tilde{\eta}\to 0$ in (\ref{qf}) and obtain
\bea\nn \mathcal{C}^q &=& c_{\Delta}^{q}\pi^{\frac{3}{2}}\frac{\Gamma\left(\frac{\Delta_q}{2}\right)}
{\Gamma\left(\frac{\Delta_q+1}{2}\right)}\frac{(-1)^q\left[2-(1+v^2)\kappa^2\right]^{q}}
{(1-v^2)^{q-1}\sqrt{\kappa^2(1-v^2 \kappa^2)}}\sum_{k=0}^{q}\frac{q!}{k!(q-k)!}
\\ \nn &&  \times \left[-\frac{1-v^2\kappa^2}{1-\frac{1}{2}(1+v^2)\kappa^2}\right]^k
{}_2 F_1 \left(\frac{1}{2},\frac{1}{2}-k;1;\frac{(1-v^2)\kappa^2}
{1-v^2\kappa^2}\right).\eea
This is exactly what was found in \cite{B1108} for finite-size giant magnons with one nonzero angular momentum.

Let us consider two particular cases. From (\ref{qf}) it follows that the normalized structure constants
for the first two string levels, for the case at hand, are given by

$q=1$ (level $n=0$)

\bea\nn \mathcal{C}^1_\eta &=& 2 c_{\Delta}^{1}\pi^{\frac{1}{2}}\frac{\Gamma\left(\frac{\Delta_1^\eta}{2}\right)}
{\Gamma\left(\frac{\Delta_1^\eta+1}{2}\right)}\frac{1}
{\sqrt{\kappa^2(1+\tilde{\eta}^2 \kappa^2)(1-v^2 \kappa^2)}}\times
\\ \nn && \Bigg[\pi(1-v^2 \kappa^2) F_1 \left(\frac{1}{2},1,-\frac{1}{2};1;\frac{\tilde{\eta}^2(1-v^2)\kappa^2}
{1+\tilde{\eta}^2\kappa^2},\frac{(1+\tilde{\eta}^2)(1-v^2)\kappa^2}
{(1+\tilde{\eta}^2\kappa^2)(1-v^2\kappa^2)}\right)
\\ \nn &&-\left(2-(1+v^2)\kappa^2\right) \mathbf{K}\left(\frac{(1+\tilde{\eta}^2)(1-v^2)\kappa^2}
{(1+\tilde{\eta}^2\kappa^2)(1-v^2\kappa^2)}\right)\Bigg].\eea

$q=2$ (level $n=1$)

\bea\nn \mathcal{C}^2_\eta &=& 2 c_{\Delta}^{2}\pi^{\frac{3}{2}}\frac{\Gamma\left(\frac{\Delta_2^\eta}{2}\right)}
{\Gamma\left(\frac{\Delta_2^\eta+1}{2}\right)}\frac{\left(2-(1+v^2)\kappa^2\right)^2}
{(1-v^2)\sqrt{\kappa^2(1+\tilde{\eta}^2 \kappa^2)(1-v^2 \kappa^2)}}\times
\\ \nn &&\Bigg\{ \frac{1}{\pi} \mathbf{K}\left(\frac{(1+\tilde{\eta}^2)(1-v^2)\kappa^2}
{(1+\tilde{\eta}^2\kappa^2)(1-v^2\kappa^2)}\right)-\frac{2(1-v^2 \kappa^2)}{\left(2-(1+v^2)\kappa^2\right)^2}
\times
\\ \nn &&\Bigg[\left(2-(1+v^2)\kappa^2\right) F_1 \left(\frac{1}{2},1,-\frac{1}{2};1;\frac{\tilde{\eta}^2(1-v^2)\kappa^2}
{1+\tilde{\eta}^2\kappa^2},\frac{(1+\tilde{\eta}^2)(1-v^2)\kappa^2}
{(1+\tilde{\eta}^2\kappa^2)(1-v^2\kappa^2)}\right)
\\ \nn &&-(1-v^2\kappa^2)F_1 \left(\frac{1}{2},2,-\frac{3}{2};1;\frac{\tilde{\eta}^2(1-v^2)\kappa^2}
{1+\tilde{\eta}^2\kappa^2},\frac{(1+\tilde{\eta}^2)(1-v^2)\kappa^2}
{(1+\tilde{\eta}^2\kappa^2)(1-v^2\kappa^2)}\right)\Bigg] \Bigg\} .\eea

In the limit $\tilde{\eta}\to 0$, the above two expressions simplify to
\bea\nn \mathcal{C}^1 &=& 2 c_{\Delta}^{1}\pi^{\frac{1}{2}}\frac{\Gamma\left(\frac{\Delta_1}{2}\right)}
{\Gamma\left(\frac{\Delta_1+1}{2}\right)}\frac{1}
{\sqrt{\kappa^2(1-v^2 \kappa^2)}}\times
\\ \nn && \Bigg[2(1-v^2 \kappa^2) \mathbf{E}\left(\frac{(1-v^2)\kappa^2}
{1-v^2\kappa^2}\right)
\\ \nn
&&-\left(2-(1+v^2)\kappa^2\right)
\mathbf{K}\left(\frac{(1-v^2)\kappa^2}
{(1-v^2\kappa^2)}\right)\Bigg] ,\eea
and
\bea\nn \mathcal{C}^2 &=& 2 c_{\Delta}^{2}\pi^{\frac{1}{2}}\frac{\Gamma\left(\frac{\Delta_2}{2}\right)}
{\Gamma\left(\frac{\Delta_2+1}{2}\right)}\frac{1}
{(1-v^2)\sqrt{\kappa^2(1-v^2 \kappa^2)}}\times
\\ \nn &&\Bigg[\left(2-(1+v^2)\kappa^2\right)^2 \mathbf{K}\left(\frac{(1-v^2)\kappa^2}
{(1-v^2\kappa^2)}\right)
\\ \nn && -4\left(2-(1+v^2)\kappa^2\right)(1-v^2\kappa^2) \mathbf{E}\left(\frac{(1-v^2)\kappa^2}
{1-v^2\kappa^2}\right)
\\ \nn && +2\pi (1-v^2\kappa^2)^2 {}_2 F_1\left(\frac{1}{2},-\frac{3}{2};1;\frac{(1-v^2)\kappa^2}
{1-v^2\kappa^2}\right)\Bigg] .\eea
respectively.

\setcounter{equation}{0}
\section{Concluding Remarks}
In this paper, in the framework of the semiclassical approach, we computed the normalized structure constants
for some three-point correlation functions in $\eta$-deformed $AdS_5\times S^5$.
Namely, we found the normalized structure constants in several classes of three-point correlators.
This was done for the cases when the ``heavy'' string states are finite-size giant magnons carrying one angular momentum
and for three different choices of the ``light'' state:
\begin{enumerate}
\item{Primary scalar operators}
\item{Dilaton operator with
nonzero momentum}
\item{Singlet scalar operators on higher string
levels}

\end{enumerate}
The results are given in terms of hypergeometric functions of two variables.
In the limit $\tilde{\eta}\to 0$, when the deformation disappear,
they reduce to the known results for the undeformed case.

%Let us mention that in \cite{LRT1411} it was established that for
%the $\eta$-deformed $AdS_2\times S^2$ subspace, the deformed
%metric can be extended to a full supergravity solution with
%non-trivial dilaton, RR scalar and RR 5-form strength.

%Since the giant magnons live in the $R_t\times S^2_{\eta}$
%subspace of the $\eta$-deformed $AdS_2\times S^2$, we believe that
%our results are correct (see \cite{LRT1411}).

It is interesting to know the explicit dependence of the
considered structure constants on the conserved angular momentum
$J_1$ of the string. However, it is not easy to find the answer to
this question. This problem will be considered in an another
paper. What is known by now, can be found in \cite{B1212} for the
$AdS_5\times S^5$ and in \cite{B1304} for the $TsT$-deformed
$AdS_5\times S^5$.

A natural generalization of our considerations here will be to
consider the case of {\it dyonic} giant magnons with two nonzero
angular momenta. We hope to report on this soon.

\section*{Acknowledgements}
This work is partially supported by the NSF grant DFNI T02/6.

\section*{Appendix}
\def\theequation{A.\arabic{equation}}
\setcounter{equation}{0}
\begin{appendix}

Here we provide the computations needed for obtaining the explicit
expression for $\mathcal{C}_{\eta}^{pr,j}$.

The integration over $\sigma$ in (\ref{c3pri}) can be changed in
the following way \bea\label{int}\int_{-L}^{L} d\sigma
=2\int_{\chi_m}^{\chi_p}\frac{d\chi}{\chi^{'}},\eea where \bea\nn
\chi_m=\chi_{min},\h \chi_p=\chi_{max},\eea and according to
\cite{AP2014} \bea\label{cp} &&\chi'=\frac{2\tilde{\eta}}{1-v^2}
\sqrt{(\chi_{\eta}-\chi)(\chi_p-\chi)(\chi-\chi_m)\chi}\ ,
\\ \label{pars}
&&\tilde{\eta}=\frac{2\eta}{1-\eta^{2}},\h
\chi_{\eta}=1+\frac{1}{\tilde{\eta}^2},\h \chi_p=1-v^2 \kappa^2,\h
\chi_m=1-\kappa^2.\eea Therefore, the normalized structure
constant can be represented as \bea\label{c3prs}
\mathcal{C}_{\eta}^{pr,j}&=&\frac{2
c_{\Delta}^{pr,j}\sqrt{\pi}}{\tilde{\eta}\kappa} \frac{\Gamma
(\frac{j}{2})}{\Gamma (\frac{1+j}{2})}
 \left[\frac{1-\kappa^{2}+j(1-v^2\kappa^{2})}{1+j} J_j-J_{jp}\right], \eea
 where
\bea\label{Jj} J_j=\int_{\chi_m}^{\chi_p}
\frac{\chi^{\frac{j}{2}}}
{\sqrt{(\chi_\eta-\chi)(\chi_p-\chi)(\chi-\chi_m)\chi}}\
d\chi,\eea \bea\label{Jjp} J_{jp}=\int_{\chi_m}^{\chi_p}
\frac{\chi^{\frac{j}{2}+1}}
{\sqrt{(\chi_\eta-\chi)(\chi_p-\chi)(\chi-\chi_m)\chi}}\ d\chi
.\eea

To compute the above two integrals, we introduce the variable
\bea\nn x=\frac{\chi-\chi_m}{\chi_p-\chi_m} \in (0,1).\eea Then
$J_j$ becomes \bea\label{Jj1}
J_j=\chi_m^{\frac{j-1}{2}}(\chi_\eta-\chi_m)^{-\frac{1}{2}}
\int_{0}^{1}
x^{-\frac{1}{2}}(1-x)^{-\frac{1}{2}}\left(1-\frac{\chi_p-\chi_m}
{\chi_\eta-\chi_m}x\right)^{-\frac{1}{2}}
\left(1+\frac{\chi_p-\chi_m} {\chi_m}x\right)^{\frac{j-1}{2}}
dx.\eea Comparing the above expression with the integral
representation for the hypergeometric function of two variables
$F_1(a,b_1,b_2;c;z_1,z_2)$ \cite{PBMv3} \bea\nn
F_1(a,b_1,b_2;c;z_1,z_2)= \frac{\Gamma(c)}{\Gamma (a)\Gamma(c-a)}
\int_{0}^{1} x^{a-1}(1-x)^{c-a-1}(1-z_1 x)^{-b_1}(1-z_2 x)^{-b_2},
\\ \nn Re(a)>0,\h Re(c-a)>0,\eea
one finds \bea\label{Jj2}
J_j=\pi\chi_m^{\frac{j-1}{2}}(\chi_\eta-\chi_m)^{-\frac{1}{2}} F_1
\left(\frac{1}{2},\frac{1}{2},-\frac{j-1}{2};1;\frac{\chi_p-\chi_m}
{\chi_\eta-\chi_m},-\frac{\chi_p-\chi_m} {\chi_m}\right).\eea

In order to compute $J_{jp}$, we have to replace $j$ with $j+2$.
Doing this, we obtain \bea\label{Jjp2}
J_{jp}=\pi\chi_m^{\frac{j+1}{2}}(\chi_\eta-\chi_m)^{-\frac{1}{2}}
F_1
\left(\frac{1}{2},\frac{1}{2},-\frac{j+1}{2};1;\frac{\chi_p-\chi_m}
{\chi_\eta-\chi_m},-\frac{\chi_p-\chi_m} {\chi_m}\right).\eea

The replacement of (\ref{Jj2}) and (\ref{Jjp2}) into (\ref{c3prs})
gives \bea\label{c3prsr}
\mathcal{C}_{\eta}^{pr,j}&=&\frac{2\pi^{\frac{3}{2}}c_{\Delta}^{pr,j}}{\tilde{\eta}\kappa}
\frac{\Gamma (\frac{j}{2})}{\Gamma (\frac{1+j}{2})}
\frac{\chi_m^{\frac{j-1}{2}}}{\sqrt{\chi_\eta-\chi_m}}
\Bigg\{\left[1-\frac{(1+jv^2)\kappa^{2}}{1+j}\right]\times
\\ \nn &&F_1 \left(\frac{1}{2},\frac{1}{2},\frac{1-j}{2};1;\frac{\chi_p-\chi_m}
{\chi_\eta-\chi_m},-\frac{\chi_p-\chi_m} {\chi_m}\right)
\\ \nn &&-\chi_m F_1 \left(\frac{1}{2},\frac{1}{2},-\frac{1+j}{2};1;\frac{\chi_p-\chi_m}
{\chi_\eta-\chi_m},-\frac{\chi_p-\chi_m} {\chi_m}\right)\Bigg\}.
\eea

Knowing that according to (\ref{pars}) \bea\nn
\chi_\eta=1+\frac{1}{\tilde{\eta}^{2}}\h \chi_p=1-v^2 \kappa^2,\h
\chi_m=1-\kappa^2,\eea and using the relation \cite{PBMv3} \bea\nn
F_1(a,b_1,b_2;c;z_1,z_2)=(1-z_1)^{c-a-b_1}(1-z_2)^{-b_2} F_1
\left(c-a,c-b_1-b_2,b_2;c;z_1,\frac{z_1-z_2}{1-z_2}\right),\eea we
can rewrite (\ref{c3prsr}) in the following form
\bea\label{c3prsfA}
\mathcal{C}_{\eta}^{pr,j}&=&\frac{2\pi^{\frac{3}{2}}c_{\Delta}^{pr,j}\Gamma
(\frac{j}{2}) (1-v^2 \kappa^2)^{\frac{j-1}{2}}} {\Gamma
(\frac{1+j}{2})\sqrt{\kappa^2(1+\tilde{\eta}^2 \kappa^2)}}
\Bigg\{\left[1-\frac{(1+jv^2)\kappa^{2}}{1+j}\right]\times
\\ \nn &&F_1 \left(\frac{1}{2},\frac{j}{2},\frac{1-j}{2};1;\frac{\tilde{\eta}^2(1-v^2)\kappa^2}
{1+\tilde{\eta}^2\kappa^2},\frac{(1+\tilde{\eta}^2)(1-v^2)\kappa^2}
{(1+\tilde{\eta}^2\kappa^2)(1-v^2\kappa^2)}\right)
\\ \nn &&-(1-v^2\kappa^2) F_1 \left(\frac{1}{2},\frac{2+j}{2},-\frac{1+j}{2};1;\frac{\tilde{\eta}^2(1-v^2)\kappa^2}
{1+\tilde{\eta}^2\kappa^2},\frac{(1+\tilde{\eta}^2)(1-v^2)\kappa^2}
{(1+\tilde{\eta}^2\kappa^2)(1-v^2\kappa^2)}\right)\Bigg\} .\eea
This is our final exact semiclassical result for this type of
three-point correlation functions.

\end{appendix}

\end{document}